# Ground Truth Bias in AI-Assisted Medical Annotation: A Crossover Study Using an Independent Radiologist Reference Standard

**Running title:** Ground truth bias and AI-assisted annotation

Márton Baltay[1]*, Tamás Ilcsik[2], Márton Borbély MD[3]

**Affiliations**

[1]Department of Sociology, Faculty of Humanities and Social Sciences, Pázmány Péter Catholic University, Budapest, Hungary

[2]School of Computing, Engineering and Physical Sciences, University of the West of Scotland, Lanarkshire, United Kingdom

[3]Department of Integrated Radiology, North Buda Saint John's Centre Hospital, Budapest, Hungary

Corresponding author: Márton Baltay, Fejér György utca 12, 2094 Nagykovácsi, Pest County, Hungary, baltay.marton.mihaly@hallgato.ppke.hu, +36 20 397 1062, https://orcid.org/0009-0005-4484-2597

## Abstract

Behavioural evaluations of artificial-intelligence-assisted annotation score every human decision against a reference standard, so their conclusions inherit that standard's validity. We tested how far conclusions about assistance, reliance and individual differences depend on it. Fifty-one non-expert high-school students annotated 50 knee radiographs for osteoarthritis, with and without a deliberately imperfect classifier, in a crossover design; personality and non-verbal reasoning were also measured. Decisions were scored against the public repository labels and against an independent reference standard from a radiologist panel, which graded 23 of the 50 images as indeterminate. Against the repository labels, assistance raised accuracy from 60.4% to 65.3% (odds ratio 1.23, p=0.002). Against the independent standard, the gain on the 27 determinate images was smaller and not significant in the primary model (78.4% to 80.8%; odds ratio 1.16, p=0.178). The apparent benefit had been concentrated on the indeterminate images and was not significant when those images were assigned to either class. The direction of residual over-reliance depended on how indeterminate cases were handled, and the personality trait most strongly associated with accuracy changed identity and sign between standards. Reliance measures are only as valid as the reference standard that defines when the assistant is right.

## Introduction

Medical image annotation remains a major bottleneck in clinical artificial intelligence (AI) development. Because expert annotation is costly and slow, pipelines utilizing non-expert annotators supported by AI predictions, training materials, and visual explanations have become an attractive operational alternative, shifting the labeling burden from scarce clinicians while preserving label quality [1-3].

However, this operational strategy introduces a methodological issue. AI-assisted annotation studies evaluate human-AI collaboration: whether humans accept correct advice, reject errors, or succumb to automation bias. This behavioral assessment depends entirely on reference-standard validity. If the reference standard is biased, evaluations may measure how human and model errors align with historical dataset inaccuracies rather than genuine human judgment.

This issue is compounded when the same public dataset is used for model training and human evaluation. If the repository contains systematic false negatives, the trained model will reproduce these errors [4,5]. A participant who agrees with an AI false negative is scored as correct because the model and reference standard align in error, while a participant who correctly identifies pathology missed by both is penalized. Consequently, AI benefits are inflated, over-reliance is hidden, and traits predicting appropriate skepticism appear ineffective.

This links our work to multi-reader multi-case (MRMC) reader studies, which are designed to separate reader, case, and modality effects [6-8]. Although not a conventional diagnostic trial, our study shares this structural vulnerability: the apparent effect of AI assistance remains dependent on case-level reference-standard validity.

Additionally, noisy-label research shows that simple accuracy metrics obscure clinically asymmetric harms. Since false negatives and false positives carry different consequences, risk-aware frameworks must evaluate directional errors and asymmetries [9-13]. We extend this logic to behavioral evaluations: the critical issue is whether the direction of ground truth noise systematically rewards unsafe reliance.

Knee osteoarthritis grading using the Kellgren-Lawrence (KL) scale is particularly vulnerable due to subjective boundaries. Distinguishing KL0, KL1, and KL2 relies on subtle osteophytes and joint-space narrowing, which exhibit low inter-rater agreement [14-18].

Automation bias leads users to substitute AI outputs for independent judgment, causing omission errors (overlooking problems the AI missed) and commission errors (accepting false suggestions) [19-23]. False-negative over-reliance is critical in imaging because the absence of an alert lacks a salient cue for review [24,25]. Cognitive forcing strategies—like requiring independent decisions before showing AI predictions—aim to promote active engagement [26-28].

Individual traits can also modulate reliance. Neuroticism is associated with uncertainty sensitivity and heightened error monitoring [29,30]. Dual-process accounts suggest that AI recommendations serve as heuristic cues; individuals high in Neuroticism may reject these cues when ambiguous [31,32]. Trait Activation Theory suggests these differences emerge under uncertainty [33-36]. However, this behavioral signal can only be observed if the reference standard rewards valid skepticism. If the ground truth shares the AI's false-negative bias, skeptical participants will be incorrectly penalized.

This study addresses two questions: (1) does AI assistance improve annotation accuracy when evaluated against an independent expert reference standard, and (2) do personality traits predict collaboration, or are these relationships suppressed by ground truth error? We analyzed a crossover experiment under both the original OAI labels and an expert radiologist consensus, testing whether apparent behavioral benefits are stable when the reference standard is scrutinized. The question matters for human-computer interaction research because the reliance categories used to evaluate human-AI collaboration, from appropriate reliance to over-reliance, are defined by whether the AI was correct, and therefore by the reference standard that decides it [27,28,48,49].

## Materials and Methods

### *Study design*

We used a crossover repeated-measures design with a multi-reader, multi-case stimulus structure. Fifty-one participants completed two annotation sessions separated by a mandatory 28-day washout interval. Participants were randomized to one of two order groups. Group A completed the no-AI condition first and the AI-assisted condition second. Group B completed the AI-assisted condition first and the no-AI condition second. This structure allowed each participant to serve as their own control while distributing condition order across participants (Figure 1). A GEE interaction model including condition and session found no statistically detectable carryover effect (condition by session p=0.554).

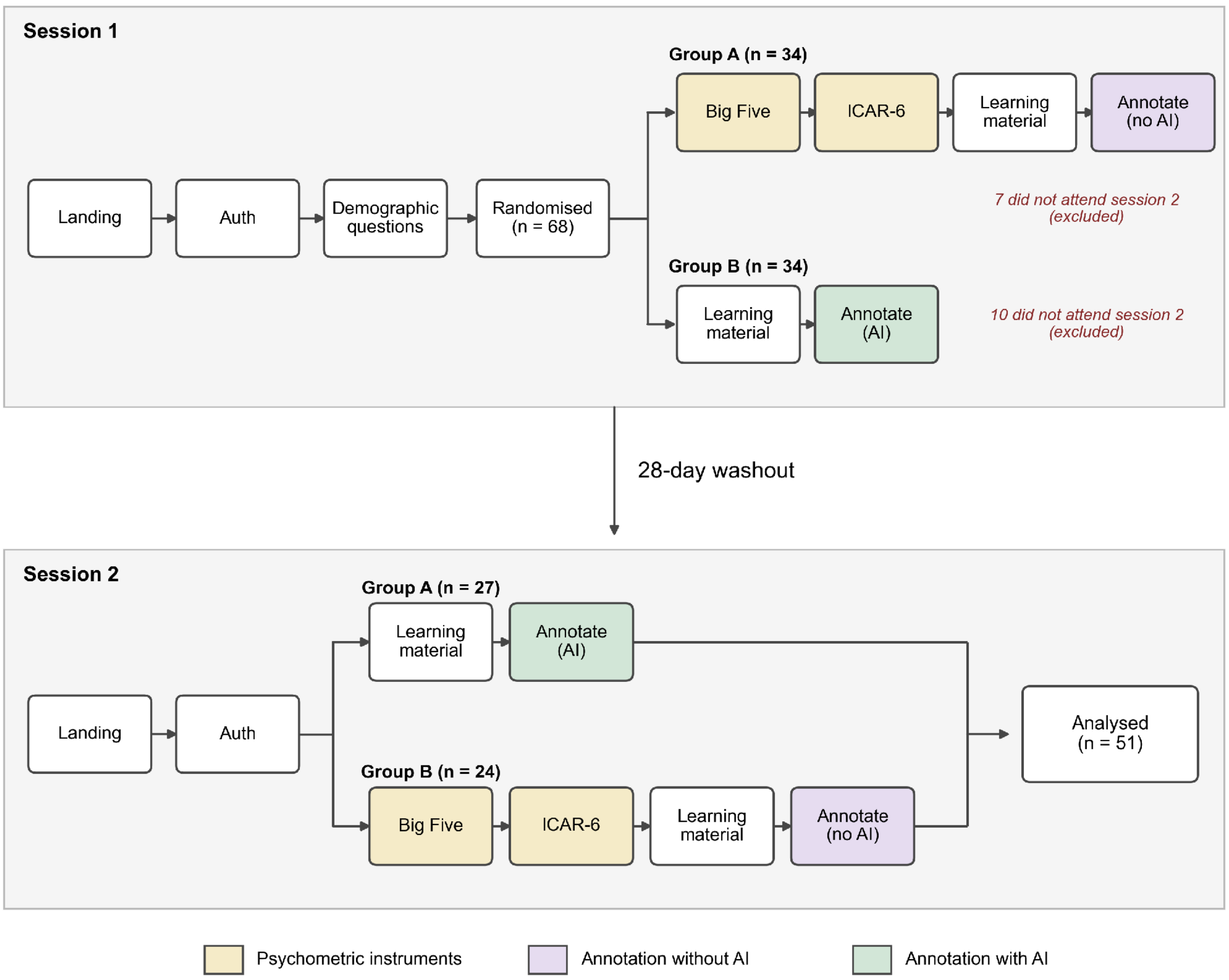

**Figure 1.** Crossover study flow diagram showing randomization (n = 34 per sequence), session order, psychometric assessment, annotation conditions, the 28-day washout, and attrition (7 and 10 participants did not attend session 2), leaving 27 and 24 completers per sequence (n = 51 analysed).

*Alt text: A flowchart depicting the repeated-measures crossover study design, showing 68 enrolled participants randomized into two groups of 34, each completing either an AI-assisted or no-AI annotation session first, a 28-day washout period, and then the alternate condition, with psychometric instruments administered at the start of each participant's no-AI session. Seven participants in the no-AI-first group and ten in the AI-first group did not attend session 2, leaving 27 and 24 completers.*

### *Participants*

Sixty-eight Hungarian high school students were recruited through school partnerships. Of these, 51 completed both annotation sessions and formed the analytic cohort. The 17 non-completers were excluded for documented session absence due to illness or scheduling conflict; 27 completers had been allocated to the no-AI-first sequence and 24 to the AI-first sequence. No participant was excluded after completing both sessions. Psychometric data were available for seven of the 17 non-completers, all from the no-AI-first sequence (the AI-first sequence completed psychometrics only in session 2); on that subset, completers and non-completers did not differ significantly on any Big Five domain, ICAR-6, age, or sex (all $p>0.09$), but the comparison is underpowered and non-differential attrition is treated as a limitation. The final cohort comprised 27 males and 24 females aged 15 to 18 years (mean 16.41, SD 0.67). No participant had medical or healthcare training, consistent with the non-expert annotation paradigm.

### *Psychometric instruments*

Participants completed psychometric instruments during their no-AI session to balance overall session duration. Big Five personality traits were measured using the 60-item BFI-2, administered in Hungarian, an instrument producing domain scores for Open-Mindedness, Conscientiousness, Extraversion, Agreeableness, and Neuroticism [37]. The cohort means were Open-Mindedness 3.77 (SD 0.63), Conscientiousness 3.38 (SD 0.72), Extraversion 3.41 (SD 0.70), Agreeableness 3.65 (SD 0.51), and Neuroticism 2.98 (SD 0.80). Non-verbal fluid intelligence was measured using the ICAR-6 matrix reasoning scale [38]. The ICAR-6 yielded a mean score of 1.12 out of 6 (SD 1.63), which falls far below normative expectations and is unlikely to reflect the cohort's true fluid intelligence. Behavioural observations during data collection — including observable hesitation and spontaneous verbal comments from participants — indicated that many students had difficulty understanding the item format rather than the reasoning content itself. The ICAR-6's abstract matrix format, presented without an example trial or guided practice, may have imposed an additional task-comprehension demand that was not adequately scaffolded for this age group in the context of an already-complex experimental session. Accordingly, the low scores are more plausibly attributed to measurement invalidity arising from inadequate task comprehension than to a genuine ability floor, and no valid inference about cognitive ability is drawn from these data. Variance inflation factors across predictors were below 1.5, indicating negligible multicollinearity.

### *Stimulus set and AI assistant*

The stimulus set consisted of 50 anteroposterior knee radiographs drawn from the test set of the OAI public repository [39,40]. The set was sampled to replicate the dataset of the DANNY study [3]: stratified random sampling with a fixed seed, KL1 images excluded from sampling, and the model's classifications fixed at 18 true negatives, seven false positives, 17 true positives and eight false negatives against the original labels. By the original repository labels, the set comprised 25 KL0 and 25 KL2-KL4 images (15 KL2, nine KL3, one KL4) and no KL1 images, so any image an independent reader graded KL1 necessarily departed from its original label. The AI assistant was a ResNet-18 convolutional neural network trained on a separate OAI training set using the original repository labels. For training, KL0 was mapped to negative and KL2, KL3, and KL4 were mapped to positive. KL1 was excluded from model training because the grade is inherently ambiguous and does not map cleanly to either binary class.

On the 50-image stimulus set the resulting model achieved 70.0% accuracy against the original OAI labels (35/50), with eight false negatives and seven false positives. This error profile was fixed by the sampling design and is not the model's natural error distribution on the test split. The level of performance was deliberate. The study was designed to examine reliance behaviour under a plausible but imperfect AI assistant, not to optimize algorithmic performance. A near-perfect model would provide little opportunity to observe appropriate skepticism or over-reliance, while a very poor model would likely be rejected outright. The model therefore functioned as a behavioural stimulus: credible enough to influence decisions, but fallible enough to create meaningful reliance choices.

### *Annotation interface and trial flow*

The annotation platform was a custom web application. In all trials, participants could inspect the radiograph using standard zoom controls. In the no-AI condition, each trial required a binary diagnosis followed by a seven-point confidence rating. In the AI-assisted condition, the interface separated the initial human judgement from subsequent AI exposure. Participants first selected radiographic features such as osteophytes and joint-space narrowing, provided an initial binary diagnosis, and rated confidence before seeing the AI result. The AI panel then displayed the model's binary prediction, numerical confidence score, and Grad-CAM heatmap overlaid on the radiograph. The confidence score was a margin-based certainty, $|2p-1|$ for the model's output probability p, shown as a percentage: 0% corresponds to a probability at the decision threshold of 0.5 and 100% to complete certainty. It is neither the probability that the prediction is correct nor the entropy-based uncertainty used in the DANNY study [3]. Participants could inspect the heatmap, original image, and feature-overlap view before submitting a final diagnosis and post-AI confidence rating.

This trial structure created a complete behavioural record for every AI-condition image: the pre-AI human decision, pre-AI confidence, AI prediction, AI confidence, visual explanation, final human decision, and post-AI confidence. It enabled classification of each AI-condition trial into reliance categories and allowed separation of genuine independent judgement from post-AI revision (Figure 2).

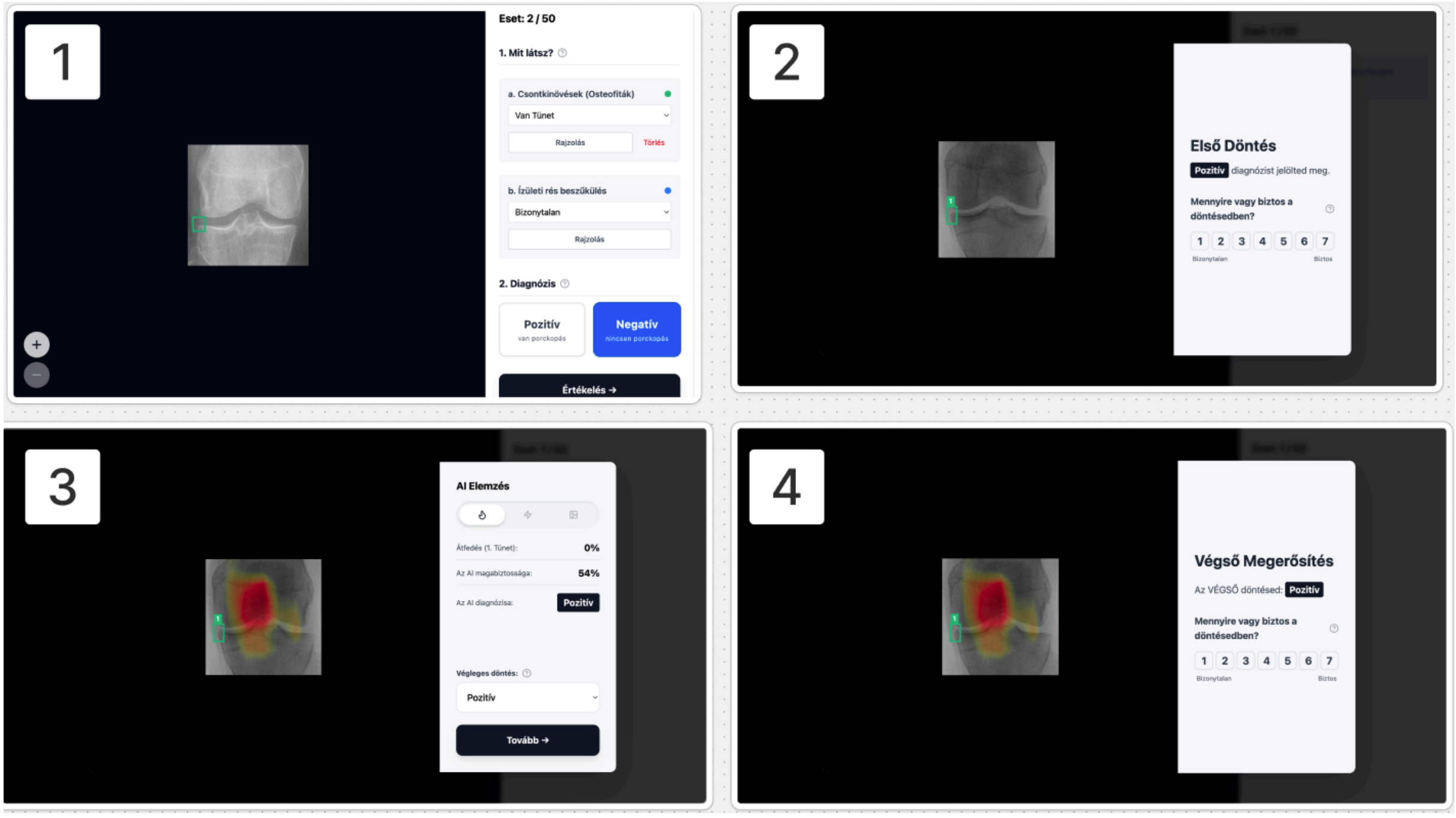

**Figure 2.** AI-condition annotation interface showing (1) feature identification and initial diagnosis, (2) initial confidence, (3) AI prediction, displayed confidence and Grad-CAM heatmap, and (4) final confirmation. Interface text is in Hungarian: Mit látsz? (What do you see?); Csontkinövések (osteophytes); Ízületi rés beszűkülés (joint-space narrowing); Első Döntés (initial decision); Az AI magabiztossága (AI confidence); Az AI diagnózisa (AI diagnosis); Végső Megerősítés (final confirmation).

*Alt text: Four numbered screenshots of the custom web application. Panel 1 shows a knee radiograph with a feature checklist and a positive or negative diagnosis choice on the right; panel 2 shows the initial-decision confirmation with a seven-point confidence scale; panel 3 shows the AI panel with the model's diagnosis, its numerical confidence and a colored Grad-CAM heatmap overlaid on the radiograph; panel 4 shows the final-decision confirmation with a second seven-point confidence scale.*

### *Independent reference standard construction*

The original OAI labels were used as the first reference standard. Preliminary analysis, however, revealed systematic disagreement between the original labels, collective non-expert annotations, and the AI model output on a subset of images. This pattern motivated an independent reference-standard validation process. One board-certified radiologist and two radiology residents independently annotated the 50 study images plus 10 additional KL1-enriched images on the full KL0-KL4 scale. The additional images were included to reduce anchoring toward a binary diagnostic frame. The three annotators were blinded to original labels, AI predictions, participant annotations, and each other's ratings during independent annotation.

The Platinum Standard was constructed through a seven-step process. First, original OAI labels were retained as the initial comparator. Second, the discordance pattern was quantified to determine whether the issue was non-random. Third, independent full-scale KL ratings were obtained from the three annotators. Fourth, binary consensus was assigned automatically where all three ratings fell on the same side of the binary boundary: KL0 as negative and KL2-KL4 as positive. Fifth, cases without complete binary consensus were reviewed during a structured consensus meeting facilitated by a third radiology resident. Sixth, inter-rater reliability was calculated using Fleiss' kappa. Seventh, all inferential models were run in parallel under both reference standards.

During independent and consensus review, adjudication was anchored to the full KL0-KL4 construct rather than to the binary label alone. The binary boundary was defined a priori as KL0 negative and KL2-KL4 positive, while KL1 was treated as diagnostically uncertain. Disagreements were resolved by discussion between the readers and were not constrained to a deterministic rule; images with identical independent rating profiles could therefore receive different consensus grades (see Discussion).

The term Platinum Standard is used here to indicate an expert consensus reference standard [41-43] constructed to provide a more defensible reference than the original public repository labels for this behavioural experiment; it is a competing expert reference, not a demonstrated improvement, and its own reliability is reported below. It does not imply that the KL scale is free from ambiguity. Rather, it acknowledges that ambiguous KL1 cases require explicit handling rather than being forced invisibly into a binary ground truth.

### *Outcome definitions and statistical analysis*

The primary outcome was binary annotation accuracy under AI-assisted versus no-AI conditions. Secondary performance metrics included balanced accuracy, F1 score, sensitivity, and specificity. Because class balance differed between the original and Platinum analyses, raw accuracy was not interpreted alone. Reliance behaviour was categorized for AI-condition trials as appropriate reliance, over-reliance, appropriate skepticism, or unwarranted skepticism. Appropriate reliance occurred when the participant agreed with a correct AI output. Over-reliance occurred when the participant agreed with an incorrect AI output. Appropriate skepticism occurred when the participant correctly overrode an incorrect AI output. Unwarranted skepticism occurred when the participant incorrectly overrode a correct AI output.

The primary inferential model was generalized estimating equations (GEE) [44] with a binomial family, logit link, and exchangeable working correlation structure. This model accounted for repeated image-level observations nested within participants. Odds ratios and 95% confidence intervals were reported. Participant-level Wilcoxon signed-rank tests were used as pre-specified sensitivity checks.

Psychometric predictors were analysed one trait at a time in a GEE model of trial-level accuracy containing the trait and the condition indicator, fitted separately under each reference standard. Entering the six predictors singly rather than jointly keeps each estimate interpretable as a marginal association and defines a single, symmetric correction family. Within each reference standard, the six resulting p-values (the five BFI-2 domains and ICAR-6) were adjusted using the Benjamini-Hochberg false discovery rate procedure; both unadjusted and adjusted values are reported throughout, and only adjusted values are used to declare significance. Trait associations with participant-level reliance rates were tested with Spearman correlations under the same correction family. Whether any trait association was specific to AI assistance was tested with a condition-by-trait interaction term.

Two aspects of the psychometric analysis were not pre-specified and are reported as exploratory. First, the study was not registered with a third-party registry before data collection. Second, the a priori directional hypothesis was that higher Neuroticism would be associated with greater over-reliance on AI advice; the observed association is in the opposite direction, so its direction was determined after inspecting the data and is treated as hypothesis-generating rather than confirmatory.

Because the primary endpoint was the binary trial-level decision, GEE was selected as the main population-averaged model; confidence-based ROC analysis was not pursued. Period (session) and sequence effects were examined by adding a session term and a condition-by-session interaction to the primary model, as recommended for crossover designs. Two further checks used the pre-AI decision recorded in every AI-condition trial: a comparison of pre-AI accuracy in the AI session with accuracy in the no-AI session, to test whether the AI-condition workflow itself altered performance, and a within-trial pre-AI versus post-AI comparison by exact McNemar test, which isolates the effect of seeing the AI output from all between-session influences. Because no attention screen was applied, a sensitivity analysis excluding no-AI trials completed in under three seconds was also run.

The Platinum Standard classified 23 of the 50 study images as KL1. Because KL1 does not map unambiguously to a binary diagnosis, three KL1 handling strategies were defined. Strategy A excluded KL1 cases from binary behavioural analysis and served as the primary analysis. Strategy B mapped KL1 to negative, reflecting conservative clinical convention. Strategy C mapped KL1 to positive, testing the upper boundary of pathology inclusion. All central GEE models were rerun under each strategy to evaluate robustness.

## Results

### *Ground truth validation*

The transition from original OAI labels to the independent radiologist reference standard moved labels toward the indeterminate middle of the scale in both directions. Of the 25 images originally labelled KL0, 14 were adjudicated as KL1 and five as KL2 or above; of the 25 originally labelled KL2 or above, nine were adjudicated as KL1 and none as KL0. Confirmed corrections that crossed the binary boundary were therefore all in the false-negative direction (five versus zero), but the asymmetry did not reach significance on a two-sided sign test (p=0.063), and the comparison is structurally unbalanced because all 25 original negatives sat adjacent to the boundary whereas ten of the original positives were KL3 or KL4. The panel also graded higher overall than the repository (mean KL 1.43 across readers versus 1.22), so a calibration difference between reader panels is an alternative explanation to systematic repository error and cannot be excluded on these data.

The panel adjudicated 23 of the 50 study images as KL1, confirming that nearly half of the stimulus set lay in the diagnostically indeterminate region of the KL scale. Under the primary Strategy A analysis, these images were excluded from binary behavioural analysis, leaving 27 unambiguous images: 21 positive and six negative. This produced a 3.5:1 class imbalance and a naive majority-class baseline of 77.8%. Pre-consensus inter-rater reliability was Fleiss' kappa=0.181 on the full five-grade scale and kappa=0.381 on binary classification across all 50 images; on the 27 images retained under Strategy A, binary kappa was 0.517, but that value is conditioned on the exclusion of the cases the panel found hardest and should not be read as the panel's overall reliability. Within-one-grade agreement was 84.0%, indicating substantial coarse agreement even where exact KL grade differed (Figure 3).

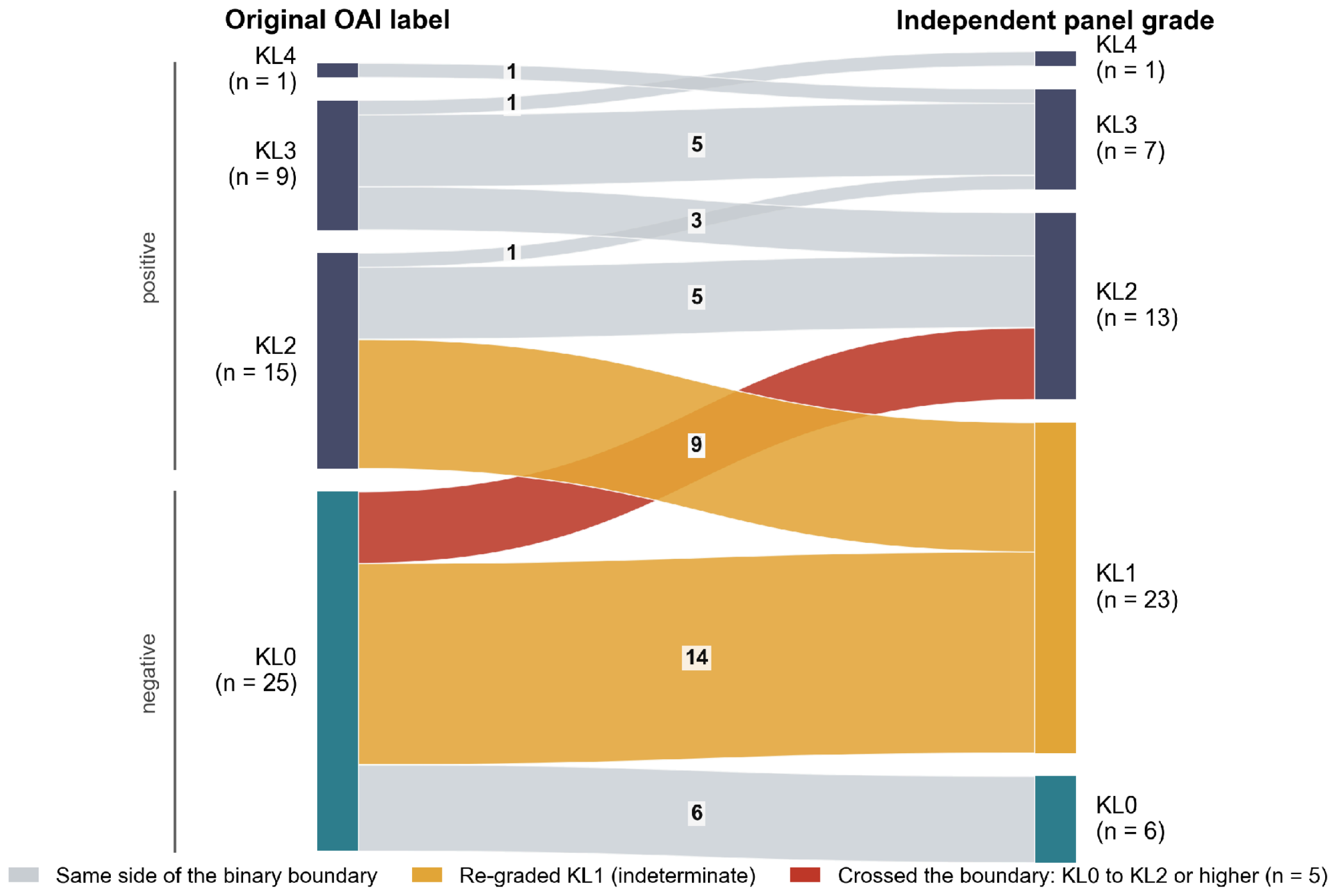

**Figure 3.** Original versus independent reference-standard label flow across Kellgren-Lawrence grades.

*Alt text: An alluvial or Sankey diagram showing the flow of knee radiographs from their original OAI Kellgren-Lawrence grades on the left to the independent panel's grades on the right, with the number of images on each flow. Of 25 original KL0 images, 14 became KL1 and five became KL2; of 15 original KL2 images, nine became KL1.*

On the five images reclassified from original KL0 to KL2 or above, the model itself predicted positive on three; that is, it disagreed with the original label on the majority of the cases on which the panel did.

### *Annotation performance under original and corrected reference standards*

**Table 1.** Human annotation performance by AI condition and reference standard.

| Ground truth | Condition | Accuracy | F1 | Sensitivity | Specificity | Balanced accuracy | GEE OR | 95% CI | p |
|---|---|---|---|---|---|---|---|---|---|
| Original OAI | AI | 65.3% | 0.676 | 72.4% | 58.2% | 0.653 | 1.232 | 1.083-1.401 | 0.002 |
| Original OAI | No-AI | 60.4% | 0.629 | 67.2% | 53.6% | 0.604 | Reference | - | - |
| Independent reference | AI | 80.8% | 0.867 | 80.7% | 81.4% | 0.810 | 1.164 | 0.933-1.453 | 0.178 |
| Independent reference | No-AI | 78.4% | 0.852 | 79.9% | 72.9% | 0.764 | Reference | - | - |

Under the original OAI ground truth, AI assistance appeared to yield a statistically strong improvement across the full 50-image dataset. Accuracy increased from 60.4% in the no-AI condition to 65.3% in the AI-assisted condition. The GEE model estimated OR=1.232 (95% CI 1.083-1.401, p=0.002), and the Wilcoxon signed-rank sensitivity check was also significant (p=0.003). The participant-level effect size was $d_z$=0.44 ($d_{av}$=0.60). Read in isolation, this would constitute a compelling demonstration of AI benefit.

Under the independent reference standard, the same participants annotating the same images produced a materially different inference. Because the independent standard identified 23 images as ambiguous (KL1), these were excluded from the primary binary analysis. On the remaining 27 unambiguous cases, No-AI accuracy rose to 78.4%, and AI-assisted accuracy reached 80.8%. Both figures should be read against the 77.8% majority-class baseline of the retained set; balanced accuracy, which is not inflated by class imbalance, was 76.4% without and 81.0% with AI assistance. The GEE model was non-significant (OR=1.164, 95% CI 0.933-1.453, p=0.178). The Wilcoxon sensitivity analysis was significant (p=0.037), indicating a small participant-level directional effect. The divergence between tests is informative: the corrected ground truth does not eliminate every possible AI benefit, but it reduces the apparent effect from a consistent improvement to a smaller and more heterogeneous signal.

The KL1 sensitivity analyses supported this interpretation. When KL1 was mapped to negative, the AI condition effect was null (OR=1.015, p=0.841). When KL1 was mapped to positive, the effect remained non-significant (OR=1.107, p=0.271). Therefore, the central conclusion did not depend on the primary exclusion of KL1 cases (Figure 4). Within Strategy A, however, the attenuation reflected which images were scored more than how they were labelled: on the same 27 images, the original labels yielded a similar and also non-significant estimate (OR=1.136, 95% CI 0.982-1.315, p=0.087), whereas on the 23 KL1 images alone they yielded a significant one (OR=1.371, 95% CI 1.144-1.643, $p<0.001$). Strategies B and C, which retain all 50 images and change only their labels, therefore provide the direct test of relabelling.

A session term added to the primary model showed a period effect under the independent standard (session 2 versus session 1, OR=0.763, p=0.008) but not under the original labels (OR=1.018, p=0.786); the period-adjusted condition estimate under the independent standard was OR=1.185 (p=0.101), and the condition-by-session interaction was not significant (p=0.554). The pre-AI decision recorded in every AI-condition trial provided two further checks. First, pre-AI accuracy in the AI session was indistinguishable from accuracy in the no-AI session (60.4% versus 60.4%, OR=0.998, p=0.976 under the original labels; 79.3% versus 78.4%, OR=1.058, p=0.606 under the independent standard), so the feature-checklist workflow of the AI condition did not by itself change performance. Second, the within-trial comparison of pre-AI with post-AI decisions reproduced the pattern of the primary analysis with greater power: under the original labels, seeing the AI output changed 184 incorrect decisions to correct and 59 correct decisions to incorrect (+4.9 percentage points, exact McNemar p<0.001); under the independent standard on the 27 unambiguous images, 61 were helped and 40 harmed (+1.5 points, p=0.046). Decomposing the original-label gain by image set showed that it was concentrated on the 23 images the panel later graded KL1 (+7.7 points, 116 helped versus 26 harmed) rather than on the 27 unambiguous images (+2.5 points, 68 versus 33). Excluding no-AI trials completed in under three seconds (479 trials, 18.8% of the no-AI arm) strengthened rather than weakened the original-label effect (OR=1.328), because fast trials were disproportionately easy cases; the primary estimates are therefore not attributable to low-effort responding.

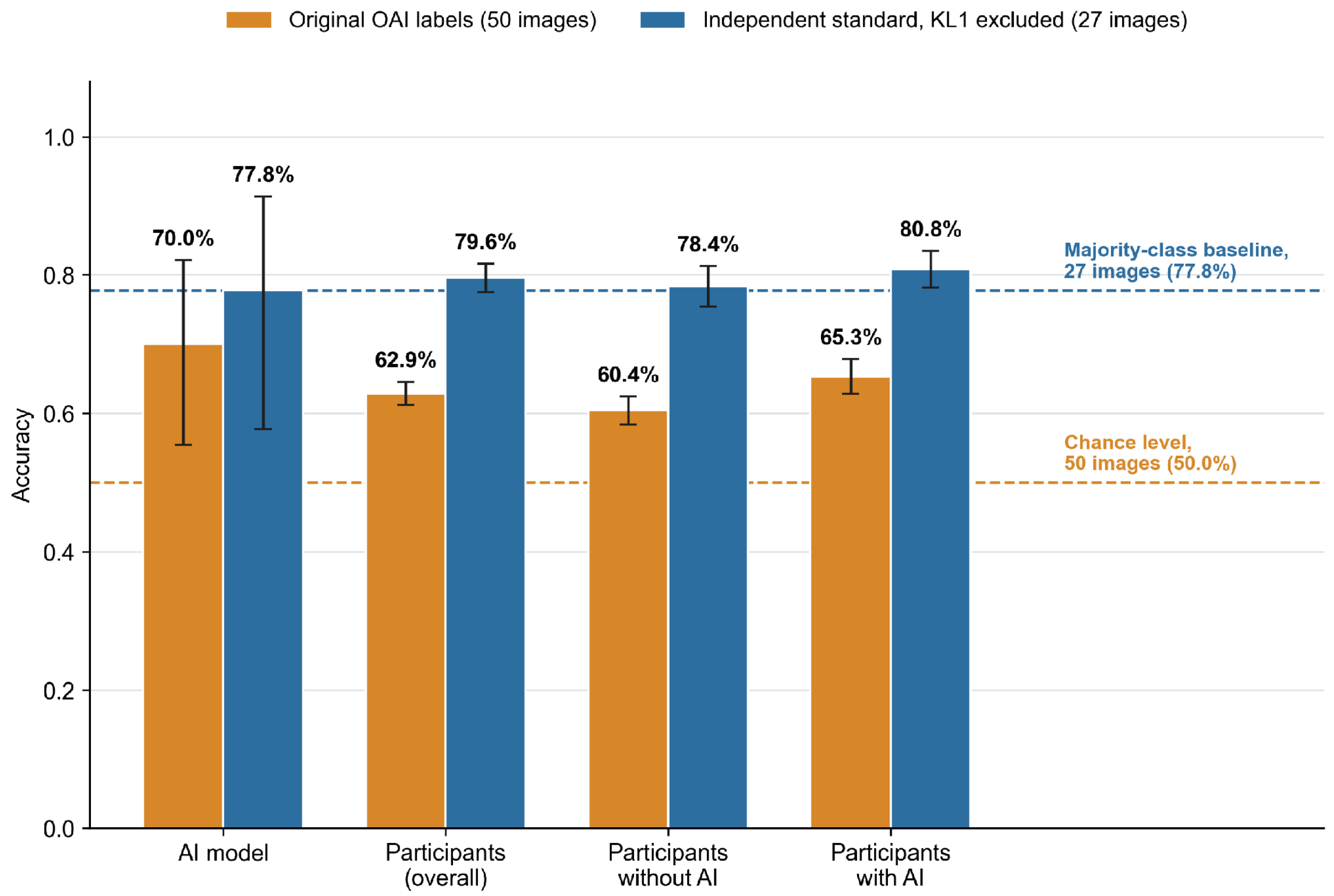

**Figure 4.** Accuracy of the AI model and of participants overall, without AI, and with AI under each reference standard. Original-label bars are computed on all 50 images and independent-standard bars on the 27 retained images. Raw accuracy rises under the corrected standard largely because Strategy A leaves a 21:6 class imbalance (majority-class baseline 77.8%, which the AI model's accuracy exactly equals); balanced accuracy, which is unaffected by this, is reported in Table 1. Dashed lines show the chance level for the 50-image set (50%) and the majority-class baseline for the 27-image set. Error bars are 95% confidence intervals: t-based across the 51 participants for human bars, and exact binomial across images for the AI model.

*Alt text: Bar charts comparing the diagnostic accuracy of the human participants, the AI model, and the AI-assisted participants under both the original OAI labels and the independent standard, with 95% confidence intervals and dashed baselines at 50% and 77.8%. All three agents show higher raw accuracy under the corrected standard, reflecting the class imbalance of the retained image set.*

### *Reliance behaviour*

**Table 2.** Reliance taxonomy by reference standard in AI-condition trials. Original OAI rates are computed across the full 50-image stimulus set; independent reference rates are computed on the 27 unambiguous cases remaining after KL1 exclusion. 95% confidence intervals are participant-clustered.

| Reliance category | Original OAI (50 images) | 95% CI | Independent reference (27 images) | 95% CI |
|---|---|---|---|---|
| Over-reliance | 15.6% | 14.5-16.6% | 7.8% | 6.1-9.4% |
| Appropriate skepticism | 14.4% | 13.4-15.5% | 14.5% | 12.9-16.1% |
| Appropriate reliance | 50.9% | 48.5-53.3% | 66.4% | 64.4-68.2% |
| Unwarranted skepticism | 19.1% | 16.7-21.5% | 11.4% | 9.5-13.3% |

The corrected reference standard changed the measured reliance taxonomy. Under the original labels applied to all 50 images, measured over-reliance was 15.6%. Under the independent reference standard applied to the 27 unambiguous cases, over-reliance fell to 7.8%. Because these rates are computed on different image sets, the reduction reflects two additive components. First, restricting the original labels to the same 27 unambiguous images reduced over-reliance from 15.6% to 11.4%, attributable to the exclusion of 23 ambiguous cases that contributed disproportionately to over-reliance events. Second, switching from the original to the corrected reference standard on those 27 images reduced over-reliance from 11.4% to 7.8% (McNemar $p<0.0001$), attributable to the reclassification of false-negative label errors. However, the most important change was directional rather than volumetric. Under the original ground truth, over-reliance appeared to include both false-positive and false-negative AI errors. Under the independent standard, every remaining over-reliance event was a false-negative acceptance: participants agreeing with the AI's failure to detect genuine pathology. This is a property of the stimulus set rather than a behavioural test result, and we report it as such. The classifier predicted negative on all six unambiguous negative cases, so no false-positive AI error existed anywhere in the Strategy A image set and no false-positive over-reliance event was available to be committed. The clinically relevant observation is not that participants preferred one error direction over another, but that the only reliance failures the corrected standard leaves visible are acceptances of missed pathology (Figure 5). This direction is specific to Strategy A. Under Strategy B, mapping KL1 to negative turns the classifier's nine positive predictions on KL1 images into false positives, and 247 of 354 over-reliance events become false-positive acceptances; under Strategy C, mapping KL1 to positive, all 533 over-reliance events are false-negative. The direction of residual over-reliance is therefore determined by the KL1 handling rule rather than by participant behaviour, and we do not treat it as a behavioural finding.

The data do not support a shared false-negative bias between the model and the original labels. The model's errors against the original labels were balanced by the sampling design (eight false negatives, seven false positives), so this set cannot show whether model and repository share such a tendency in general. Within it, on the 28 images where the panel departed from the original label the model sided with the original on 17 (61%) and against it on 11, less often than on the remaining 22 images (18, 82%), showing no preferential alignment. What changed is how positive calls were scored. Under the original labels, a positive decision on any of the 19 original-KL0 images that the panel later placed at KL1 or above was scored as an error, and three of the model's seven original-label false positives fell on the five images the panel confirmed as positive. On the 27 retained images, relabelling changed reliance categories only on the five confirmed images, and in both directions: where the model had predicted positive (three images), 101 trials of agreement moved from over-reliance to appropriate reliance, and where it had predicted negative (two images), 51 trials of agreement moved from appropriate reliance to over-reliance. The reduction in unwarranted skepticism from 19.1% to 11.4% is the complementary effect of the same rescoring, and similarly reflects both image-set restriction and reference-standard correction.

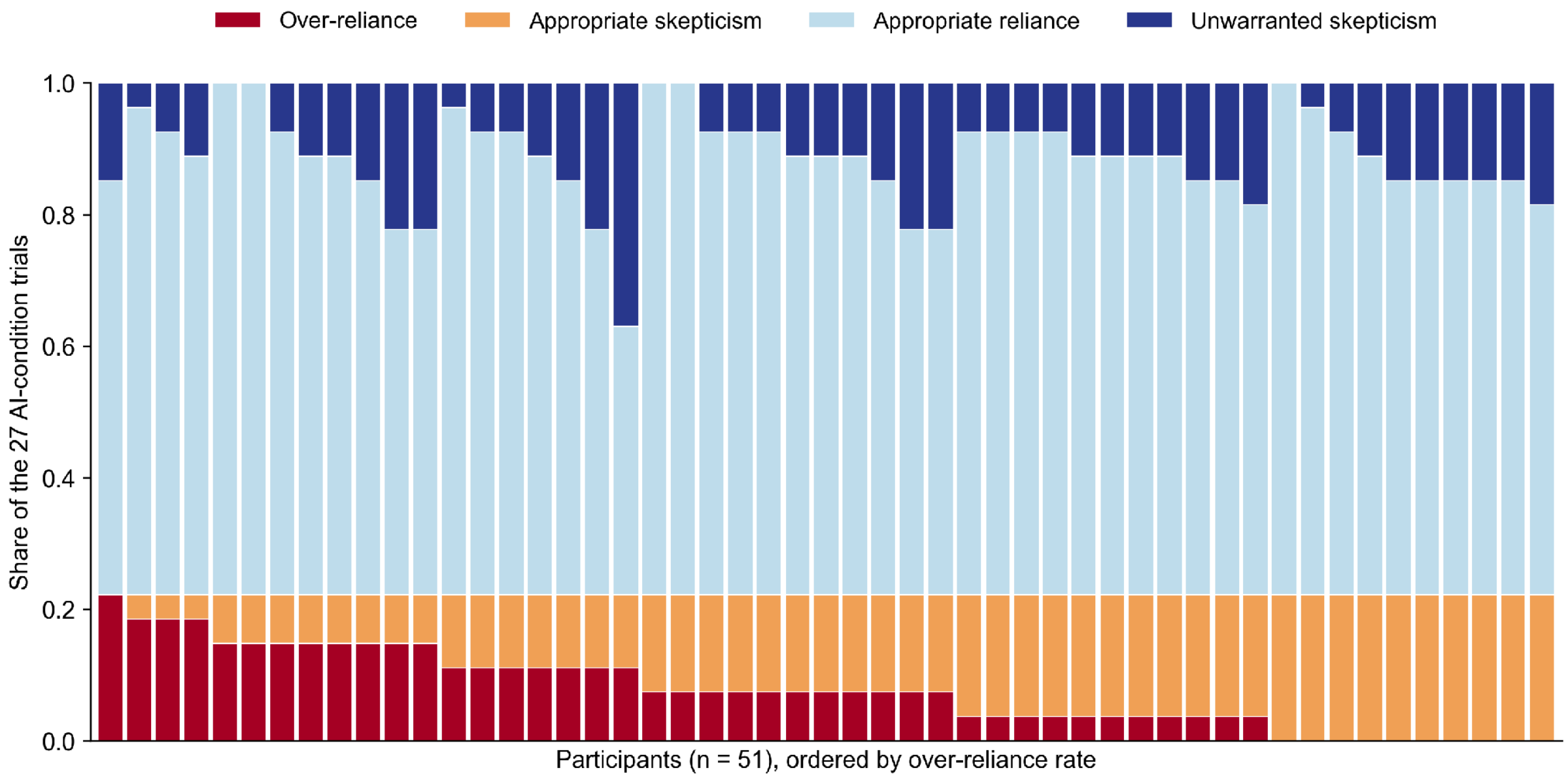

**Figure 5.** Participant-level reliance taxonomy under the independent standard (27 AI-condition trials per participant), ordered by over-reliance rate. Aggregate over-reliance is not uniform across participants.

*Alt text: A stacked bar chart showing the distribution of reliance behaviors (appropriate reliance, over-reliance, appropriate skepticism, unwarranted skepticism) for each of the 51 participants, demonstrating significant individual variance.*

### *Psychometric predictors*

**Table 3.** Trait associations with trial-level annotation accuracy under each reference standard. Each row is a separate GEE model containing the trait and the condition indicator; p_FDR is the Benjamini-Hochberg adjusted value within the six predictors tested under that reference standard. ICAR-6 estimates are reported for completeness only; the floor effect described in Methods makes them uninterpretable as cognitive-ability associations.

| Predictor | Original OAI: OR (95% CI) | p | p_FDR | Independent reference: OR (95% CI) | p | p_FDR |
|---|---|---|---|---|---|---|
| Neuroticism | 0.936 (0.857-1.022) | 0.139 | 0.294 | 1.177 (1.032-1.343) | 0.015 | 0.090 |
| Conscientiousness | 1.077 (0.974-1.191) | 0.147 | 0.294 | 0.927 (0.790-1.088) | 0.355 | 0.533 |
| Extraversion | 0.887 (0.820-0.959) | 0.003 | 0.016 | 1.131 (0.969-1.319) | 0.119 | 0.237 |
| Agreeableness | 0.987 (0.851-1.144) | 0.861 | 0.893 | 0.944 (0.721-1.236) | 0.676 | 0.676 |
| Open-Mindedness | 0.977 (0.856-1.115) | 0.731 | 0.893 | 1.068 (0.890-1.281) | 0.480 | 0.576 |
| ICAR-6 | 1.003 (0.964-1.043) | 0.893 | 0.893 | 0.949 (0.897-1.004) | 0.066 | 0.199 |

The psychometric analysis showed the same dependence on the reference standard, but in a form that is more cautionary than the accuracy results and that we deliberately do not interpret as a personality finding.

Under the original OAI labels, one trait association survived correction: Extraversion predicted lower annotation accuracy (OR=0.887, 95% CI 0.820-0.959, p=0.003, p_FDR=0.016). The direction and magnitude were stable across specifications, persisting with session adjustment, in a joint model containing all six predictors (OR=0.824, p<0.001), and at the participant level (Spearman rho=-0.379, p=0.006). Conscientiousness, which we had expected a priori to predict baseline accuracy, showed no association (OR=1.077, p_FDR=0.294).

Under the independent reference standard, that association disappeared and reversed in sign (OR=1.131, p_FDR=0.237). The largest remaining association was Neuroticism with higher accuracy (OR=1.177, 95% CI 1.032-1.343, p=0.015), but it did not survive correction for the six predictors tested (p_FDR=0.090). Neuroticism had likewise reversed sign, from OR=0.936 under the original labels to OR=1.177 under the corrected standard. No trait met the adjusted significance threshold under the independent standard. A model additionally adjusting for case grade and AI correctness returned a comparable Neuroticism estimate (OR=1.178, 95% CI 1.017-1.366, unadjusted p=0.029), which does not alter this conclusion.

One trait-reliance association did survive correction. Higher Neuroticism was associated with a lower participant-level over-reliance rate under the independent standard (Spearman rho=-0.390, 95% CI -0.603 to -0.128 by bootstrap, p=0.005, p_FDR=0.028). Two qualifications limit what can be drawn from it. First, it runs opposite to our a priori directional hypothesis, so its direction is post hoc. Second, it is not specific to AI assistance: the condition-by-Neuroticism interaction on accuracy was null (p=0.828), and the trait-accuracy correlation was in fact stronger in the no-AI condition (rho=0.312, p=0.026) than in the AI condition (rho=0.135, p=0.343). A trait that tracks accuracy at least as strongly when no AI advice is present cannot be read as an index of calibrated skepticism toward AI, and we do not read it that way here.

The pattern across Table 3 is therefore not that a genuine trait signal was suppressed by the original labels and released by the corrected ones. It is that which trait appears to predict accuracy, and in which direction, changed when the decisions of the same 51 participants were rescored against a different reference standard. With six correlated predictors, 51 participants, and two scorings of one dataset, that instability is the expected behaviour of a small-sample exploratory analysis, and it is also the clearest available illustration of this paper's central claim: reference-standard choice propagates into secondary inferences as well as primary ones. We therefore report the psychometric results as a demonstration of that fragility and draw no substantive personality conclusion from them. Confirmation would require a larger, prospectively registered sample with the trait, the direction, and the reference standard all fixed in advance.

## Discussion

This study demonstrates that the apparent benefit of AI assistance in non-expert medical annotation depends strongly on reference-standard validity. Under the original OAI labels, AI assistance produced a statistically significant improvement in accuracy. Under an independent radiologist consensus reference standard, however, the same data yielded only a small and non-significant effect in the primary GEE model. Thus, apparent AI benefits cannot be safely interpreted unless the reference standard is independently validated and its error structure understood.

The underlying mechanism is a scoring artifact of a contested reference standard rather than, in this stimulus set, a shared error bias between the model and its training labels. The model's errors against the original labels were balanced by the sampling design, so this set cannot speak to the model's natural error direction; within it, the model showed no preferential alignment with the original label on the images the panel disputed. What the original labels did was score positive calls on 19 original-KL0 images as errors, including images the panel later graded positive or indeterminate; and the assistant's apparent benefit under those labels was concentrated on precisely the 23 images the panel could not place, where it steered participants toward the repository's label. An evaluation metric that rewards agreement with a label set rewards agreement with that set's conventions and errors alike, and a model trained on the same set will appear most helpful where those conventions are least clinically defensible. This extends prior label-noise research to human-AI behavioral evaluation [9,45-47].

These findings align with radiology reader-study methodologies that emphasize reference-standard validity alongside reader-case variance. Conventional multi-reader multi-case (MRMC) frameworks account for reader and case heterogeneity, but our results show that reference-standard bias can fundamentally distort these models. Independent adjudication is not merely a procedural step; it determines the behavioral validity of reliance, skepticism, and apparent AI benefit. It does not, however, settle which of two disagreeing reference standards is closer to the truth. The independent panel's five-grade agreement was slight (kappa=0.181), its readers differed systematically in calibration, and it graded higher on average than the repository; a panel calibration difference therefore remains a competing explanation for the label migration, and the defensible claim from this design is that behavioural conclusions were unstable across two expert-derived reference standards, not that one of them was corrected.

The reliance taxonomy highlights clinical safety implications: under the corrected standard with KL1 excluded (Strategy A), all over-reliance events were false-negative acceptances (participants accepting the AI's failure to detect pathology), because the classifier made no false-positive error on the six unambiguous negatives; under Strategy B the balance reverses toward false-positive acceptances (see Results). Under the original labels, acceptance of the assistant's negative call on the two confirmed-positive images where the model also predicted negative was scored as correct, so a small number of clinically relevant reliance errors were invisible to the original evaluation. Studies that evaluate human-AI collaboration without validating reference standards risk underestimating automation bias. This underscores the need for risk-aware clinical evaluation that separates error types rather than reporting aggregate metrics [12,13].

The high proportion of KL1 cases (23/50) highlights the clinical ambiguity of early osteoarthritis. Rather than flattening these borderline cases into binary classes, researchers must handle them transparently. Our sensitivity analyses showed that the accuracy conclusion was robust across the three KL1 handling strategies, whereas the direction of residual over-reliance was not; both outcomes reinforce the necessity of reporting how ambiguous cases are managed.

The psychometric results extend the same argument to secondary inferences, and they do so more starkly than the accuracy analysis. Rescoring the identical decisions against a different reference standard changed which trait reached adjusted significance and reversed the sign of both candidate associations: Extraversion predicted lower accuracy under the original labels ($p_{FDR}=0.016$) and nothing under the corrected standard, while Neuroticism moved in the opposite direction without meeting the adjusted threshold ($p_{FDR}=0.090$). We considered the interpretation that the original labels suppressed a genuine Neuroticism effect, and rejected it: the association is not specific to AI assistance (interaction $p=0.828$) and is stronger without AI than with it, which is incompatible with a skepticism-toward-AI account. The more defensible reading is that individual-difference associations estimated on 51 participants across six correlated traits are unstable, and that reference-standard choice is one of the things they are unstable with respect to. This is a warning for a common study design rather than a personality finding, and it argues that behavioural AI-assistance studies reporting trait moderators should pre-register the trait and direction and report the reference standard's construction alongside them.

Finally, while non-expert adolescent annotators are relevant for evaluating scalable labeling pipelines, they do not substitute for clinical experts. Replicating this study with clinicians would evaluate these dynamics under realistic expertise and accountability conditions, where MRMC or mixed-effects models would be essential.

This study has several limitations. First, the adolescent cohort limits generalizability to clinical readers. Second, the stimulus set was small, with primary Strategy A analysis restricted to 27 unambiguous images. Third, the AI assistant was a deliberately moderate-accuracy model, suitable for eliciting reliance behavior but not representative of state-of-the-art diagnostic systems. Fourth, attrition could be evaluated only partially, because psychometric data were available for only seven of the 17 non-completers. Fifth, Kellgren-Lawrence grading is subjective, and the radiologist kappa confirms persistent ambiguity. Sixth, patient-level leakage risks (e.g., contralateral knees or longitudinal follow-up images) remain possible since participant IDs were not matched across splits. Seventh, confidence-interaction and calibration analyses (e.g., Brier score changes) were not performed. Eighth, complementary analyses using crossed participant and case effects or formal MRMC ROC methods were not conducted and represent a direction for future work. Ninth, the confidence displayed with each prediction was a margin-based certainty rather than a calibrated probability or the entropy measure of earlier work, and we did not test how participants interpreted it; a displayed value near 0% marks a prediction close to the decision threshold but may have been read as a very low probability that the AI was right. Tenth, consensus grades were reached by discussion rather than by a deterministic rule, and images with identical independent rating profiles sometimes received different consensus grades; the five-versus-zero count is sensitive to these calls. Eleventh, no attention screen was applied, although the speeded-trial sensitivity analysis indicated that rapid responding did not drive the result. Twelfth, for participants in the AI-first sequence, personality and cognitive measures were collected after exposure to the AI condition.

Nevertheless, the core methodological conclusion remains robust due to the within-study design: the same participants, images, AI outputs, and statistical models yielded different conclusions under different reference standards. This design isolates the reference-standard problem directly, suggesting that any human-AI evaluation relying on unvalidated public labels is vulnerable to the same scoring artifact, especially when the AI assistant is trained on the same repository.

We recommend minimum reporting standards for behavioral AI-assistance studies, including reference-standard construction, expert validation, inter-rater reliability, error directionality, and the handling of ambiguous cases. For reliance evaluations, over-reliance should be decomposed into false-negative and false-positive rates rather than aggregate metrics, and reference-standard error should be modeled as an active component of behavioral analysis.

To support reproducibility, a de-identified per-image table containing original labels, consensus labels, AI predictions, AI confidence, aggregate participant agreement rates, and reliance counts is available as supplementary material, facilitating secondary analyses within data sharing restrictions.

## Conclusion

A statistically robust AI-assistance effect can be a measurement artifact when human-AI decisions are scored against a contested public reference standard. In this crossover study, the original OAI labels suggested a significant AI benefit, while an independent radiologist reference standard reduced the effect to a small and heterogeneous signal. Under the corrected standard with KL1 excluded, all residual over-reliance was acceptance of missed pathology, and the trait associated with annotation accuracy changed identity and sign between the two standards without any trait meeting the adjusted significance threshold under the corrected one. Ground truth validation should therefore be treated as a prerequisite for behavioural inference in AI-assisted medical annotation studies, not as an optional quality check.

## Abbreviations

AI: artificial intelligence; BFI-2: Big Five Inventory-2; CI: confidence interval; FDR: false discovery rate; GEE: generalized estimating equations; Grad-CAM: Gradient-weighted Class Activation Mapping; ICAR-6: International Cognitive Ability Resource six-item scale; KL: Kellgren-Lawrence; MRMC: multi-reader multi-case; OAI: Osteoarthritis Initiative; OR: odds ratio.

## Declarations

### *Ethics approval and consent to participate*

This study was conducted in accordance with the Declaration of Helsinki and applicable Hungarian legislation. Under Hungarian law (Act CLIV of 1997 on Health), research involving minors in educational and psychological study contexts does not require formal ethics committee review when it is conducted within established school partnerships and does not involve medical procedures or clinical interventions. Accordingly, no institutional review board submission was required or sought for this non-clinical annotation study.

All participants were aged 15–18 years and provided digital informed assent prior to participation. For participants below the age of 16 (tenth-grade students), written parental or guardian consent was additionally obtained in accordance with the applicable legal requirement. Participation was fully voluntary, and all participants were informed that they could withdraw at any time without consequence. No personal health data were collected; the study recorded annotation decisions, confidence ratings, and psychometric questionnaire responses.

***Consent for publication***

Not applicable.

***Data availability statement***

Subject to OAI data-use restrictions, the authors will make the de-identified analysis code, model-output table, and per-image derived outcomes available on a public GitHub repository upon publication. The intended per-image table will include original OAI label, independent consensus label, AI prediction, AI confidence, aggregate participant agreement rates, and reliance-category counts under both reference standards, but will not include identifiable participant information or redistributable radiograph files unless permitted by the source data agreement. Participant-level data will be shared only in de-identified or aggregate form.

***Competing interests***

The authors declare that they have no competing interests.

***Funding***

No specific funding was received for this work.

***Author contributions***

Márton Baltay: Conceptualization, Data curation, Formal analysis, Methodology, Project administration, Writing – original draft, Writing – review & editing. Tamás Ilcsik: Software, Validation, Writing – review & editing. Márton Borbély: Investigation, Methodology, Project administration, Writing – review & editing.

***Acknowledgements***

The authors wish to thank László Szirmay-Kalos for his advice, Boróka Holicsné Gémes for her assistance with the student participants, and the three readers Blanka Tatár, Dorottya Perge and Péter Füssy for their independent case review and participation in the consensus meeting.

***Supplementary material***

Supplementary material comprises the de-identified per-image outcome table under both reference standards, including the three independent reader grades and the consensus grade for each image; confusion matrices under all three KL1 handling strategies; full GEE outputs for every model reported; and the analysis code.